# Title: In-plane ordering of O vacancies in a high-$T_c$ cuprate superconductor with compressed Cu-O octahedrons: a first-principles cluster expansion study

**Authors:** Yunhao Li[1,†], Shiqiao Du[2,†,*], Zheng-Yu Weng[1,*], Zheng Liu[1,2,*]

**Affiliations:**

[1]*Institute for Advanced Study, Tsinghua University, Beijing 100084, China*

[2]*State Key Laboratory of Low Dimensional Quantum Physics, Department of Physics, Tsinghua University, Beijing 100084, China*

† These authors contributed equally to this work

*Email: dsq16@mails.tsinghua.edu.cn; weng@tsinghua.edu.cn; zheng-liu@tsinghua.edu.cn

**Abstract**: A recently discovered high-$T_c$ cuprate superconductor $Ba_2CuO_{4-\delta}$ exhibits exceptional Jahn-Teller distortion, wherein the $CuO_6$ octahedrons are compressed along the *c* axis. As a consequence, the O vacancies prefer to reside in the $CuO_2$ plane, but the exact structure is not known. By combining first-principles total energy calculation with the automated structure inversion method, the effective cluster interactions of O vacancies are mapped out. Around $\delta$=0.8, where the 73K superconductivity was observed experimentally, we predict that the ordered O vacancies slice the $CuO_2$ plane into not only 1D chains and but also two-leg ladders. A Monte Carlo simulation is performed based on the effective cluster interaction model, showing that such an ordering pattern is stable up to ~900 K. Our results put forth a concrete structural basis to discuss the underlying superconducting mechanism.

**Main Text:**

High-$T_c$ superconductivity in cuprates is commonly spawned in the intact two-dimensional $CuO_2$ plane [1-6]. However, a new cuprate superconductor $Ba_2CuO_{4-\delta}$ appears to be an exception [7]. $Ba_2CuO_{4-\delta}$ crystallizes into the typical 214 layered perovskite structure [Fig. 1(a)], but the $CuO_6$ octahedrons are largely compressed along the $c$ axis, making the in-plane Cu-O bonds weaker than the out-of-plane ones. Consequently, in contrast to the apical substitution as typically observed in other high-$Tc$ cuprates [3-6], here the O vacancies prefer to be created in the $CuO_2$ plane [8,9]. Surprisingly, superconductivity emerges at a very high concentration of in-plane O vacancies, when the 2D parent lattice has been strongly disrupted, and the superconducting transition temperature reaches as high as 73 K around $\delta=0.8$.

At present, the in-plane O vacancy structure is still largely unknown except for the $\delta=1$ limit, i.e. $Ba_2CuO_3$ [8]. This stoichiometric compound as a quasi-1D Mott insulator consists of paralleled (-O-Cu-O-) chains. One can reversely consider that superconductivity emerges from $Ba_2CuO_{3+\gamma}$ around $\gamma=0.2$, when excess O atoms link the 1D chains, which appears to play an important role in the 73 K superconductivity, because in other quasi-1D cuprates with charge doping only, such a high $T_c$ has never been observed.[9-12]

An interesting observation is that the atomic structure of $Ba_2CuO_3$ can be viewed as long-range ordering of in-plane O vacancies in $Ba_2CuO_4$. $Ba_2CuO_3$ shares exactly the same lattice as $Ba_2CuO_4$, with only slight changes of the lattice constants (Tab. I). The O atoms are missing along all the unidirectional (-O-Cu-O-) arrays, like cutting either the warp or the weft of a fabric. The coordinates of the remaining atoms stay almost unchanged.

Physically, such a long-range ordering arises from specific interactions between the O vacancies. It is insightful to coarsely grain the microscopic details into a generalized Ising model [Fig .1(b)]:

$$E(\{\sigma_i\}) = \sum_\alpha J_\alpha \prod_{i \in \alpha} \sigma_i \tag{1}$$

in which at each in-plane O site, the $Z_2$ variable $\sigma_i$ is set to be $+1$ when the site is occupied by O and $-1$ if it is a vacancy. $\alpha$ denotes a specified cluster of O sites, which in general not only covers pair interactions but also multi-body interactions as dictated by the effective cluster interaction strength $J_\alpha$. In the context of alloy physics, eq. (1) is widely known as the cluster expansion (CE) method, which is able to reproduce any $E(\{\sigma_i\})$ function when all clusters α are considered in the sum.

Here, we apply the CE method to study the in-plane O vacancy structures in $BaCu_2O_4$, or reversely the excess O structures in $BaCu_2O_3$. The results will put forth a concrete structural basis to discuss the underlying superconducting mechanism. We note that a recent DFT work computed 24 O vacancy structures of $Ba_2CuO_{4-\delta}$ by first-principles calculation within the density functional theory (DFT) framework, and proposed that the lowest-energy structure featured 1D CuO chains as in $Ba_2CuO_3$. [13] By combining DFT with the CE method, we are able to enlarge the search space to more than 1800 structures. A unique two-leg ladder structure is identified that was not recognized before.

Technically, the main concern is that in constrast to substitution of chemically similar elements in alloys, a vacancy in our case is expected to induce much stronger perturbations to the local environment. It is not known a priori whether the CE converges rapidly, and likely non-compact clusters with long-range interactions have to be included, making it formidable to construct the model by hand. Fortunately, a self-learning algorithm [14] has been established to automatically decide the optimal model. Basically, the computer starts with the simplest clusters, and progressively includes more complicated ones using the existing clusters as the building blocks. For a given expansion, standard fitting of $\{J_\alpha\}$ is performed with respect to a training data set containing the total energy of a selection of ordered structures. With an increasing expansion order, the computer aims for the best compromise between underfitting and overfitting, which is achieved by minimization a target function called "Cross-Validation Score" [15].

The training data set is automatically generated by DFT calculations without human intervention, thanks to the structure enumeration algorithm [16] and the structure selection method [14]. There is no doubt that the electronic structure of cuprates is beyond the scope of DFT. Nevertheless, we find that DFT works quite well for the structural properties, which only rely on the total energy. A comparison between the experimental and the calculated parameters at the two limits $\delta = 0, 1$ are shown in Tab. I.

Figure 1(c) summarizes the CE model learned by the computer. The final cross-validation score is 0.021 eV per atom, which can be roughly viewed as a statistical estimation of the error bar of the predicted total energy. We note that this score is typically considered satisfactory in the study of alloys. The model contains 40 clusters up to quadruplets. We show the representative ones in Fig. 1(d). Overall, $\{J_\alpha\}$ decays as a function of the diameter of the corresponding cluster and as a function of the number of sites it contains [Fig. 1(c)]. Nevertheless, it is important to notice that the pair interactions have a relatively long tail, e.g. the next nearest neighbor interaction along a (-O-Cu-O-) array (p4) is still significant.

Let us first make some simple analysis when $\delta$ is close to 0 or 1. When one vacancy is added to a perfect $CuO_2$ plane [Fig. 2(a)], it just randomly picks a O site, since all are equivalent. When the second vacancy is added, it tends to form a p2 pair with the first vacancy [Fig. 2(b)], because $J_{p2}$ is a large negative number. According to Fig. 1(c), $J_{p2}$ is the only dominating attractive interaction, and all the other pair interactions as listed in Fig. 1(d) are strongly repulsive.

Let us consider a second p2 pair is added. Intuitively, we might expect that the two pairs will connect in a row to take advantage of the $J_{p2}$ attraction [Fig. 2(c)]. However, this arrangement is actually not energy favorable because of the repulsive $J_{p4}$. We can calculate the difference between $E_{4r}^v$ [Fig. 2(c)] and $2E_2^v$ [Fig. 2(b)]. The latter corresponds to the energy of two isolated p2 pairs. The result is $E_{4r}^v - 2E_2^v \approx -4|J_{p2}| + 8|J_{p4}| > 0$. The situation is similar to a frustrated spin chain, in which the nearest-neighbor interaction is ferromagnetic but the next nearest-neighbor interaction is antiferromagnetic. The latter tends to suppress the ferromagnetic order, or for our case, consecutive O vacancies in a row.

Around $\delta=1$, the same argument can be made to show that two excess atoms will form a p2 pair [Fig. 2(d)]. We arrange two separated p2 pairs in Fig. 2(e) that avoids any significant interaction listed in Fig. 1(b). Its energy $E_{4s}^O$ should be close to $2E_2^O$, apart from corrections from the long-range weak interactions. Interestingly, energy can be further reduced if we rotate the q1' quadruple between the two p2 pairs by 90 degrees [highlighted by two red circles in Figs. 2(e) and (f)]. The energy after and before the rotation has a difference: $E_{4r}^O - 2E_{4s}^O \approx 8|J_{p2}| - 16|J_{p3}| - 16|J_{p4}| < 0$. This rotation is not favored by $J_{p2}$, because it creates four O-Cu-vacancy clusters. However, much energy is gained from the repulsive $J_{p3}$ and $J_{p4}$, as indicated in Fig. 2(f). A primitive form of two-leg ladders can be readily observed after the rotation.

A systematic determination of the lowest-energy structure in the entire vacancy concentration range is numerically achieved. At a given $\delta$ between 0 and 2, the possible vacancy structures are enumerated and the energy is predicted by the CE model. The lowest-energy structure is further re-examined by DFT if it is not in the training data set. In Fig. 3(a), the green stars mark the CE energies of over 1800 different vacancy configurations. The 150 DFT training data and the corresponding CE-fitted data are highlighted by red and yellow triangles, respectively. We take $(1-\frac{\delta}{2}) \cdot E_{Ba_2CuO_4} + \frac{\delta}{2} \cdot E_{Ba_2CuO_2}$ as the reference energy at any $\delta$. In this way, the plotted energy is independent of the O chemical potential, which should be viewed as the relative energy with respect to a mixture of spatially separated $Ba_2CuO_4$ and $Ba_2CuO_2$. The lowest energy as a function of $\delta$ is overall downward convex and stays negative, suggesting that $Ba_2CuO_{4-\delta}$ does not favor phase separation.

The most important information revealed by the numerical search is that around the experimentally determined O vacancy concentration of the superconducting phase ($\delta \sim 0.8$), the lowest-energy structures typically consist of two 1D building blocks: the (-Cu-O-) chain and the three-Cu-wide two-leg ladder [Fig. 3(b)]. Recall that this ladder structure has already appeared in Fig. 2(f). We perform Monte Carlo simulation [17] to test the stability of predicted lowest-energy structure. We pick $\delta = 0.83$ as an example, and use the ensemble averaged cluster correlation functions $C_\alpha = \langle \prod_{i \in \alpha} \sigma_i \rangle$ as the phase indicator. It is straightforward to verify that the ordered structure as shown in Fig. 3(b) gives $-C_{p1} = C_{p2} = C_{p3} = 2/3$, and $C_{p4} = 1/3$. As shown in Fig. 3(c), these correlation functions are not affected until the temperature is higher than 900K.

Lastly, we briefly point out possible superconductivity inferred from the present ladder structure around $\delta = 0.8$, where the actual material shows a high-$T_c$ at 73K. As indicated in Fig. 3(b), besides the 1D chains [13] in each CuO layer, the prominent feature is another unique ladder structure composed of two Cu-O chains with yet another Cu-only chain sandwiched in between. Due to the missing O atoms along the middle Cu chain, the electron hopping along such a Cu chain or to its two neighboring CuO chains may be greatly suppressed (due to the absence of the Zhang-Rice singlet [18]). Figure 4(a) displays the calculated DFT bands of the $\delta = 0.67$ lowest-energy structure, which purely consists of the ladders. By projecting the wavefunction to atomic orbitals, the nearly flat band around 1.3 eV can be clearly attributed to the Cu in the middle (Cu2). Two dispersive bands arising from Cu on the ladder legs (Cu1) cross the flat band, extending down to the Fermi level. The dispersion is similar to that of Cu-O chains in $Ba_2CuO_3$ [13], confirming that the coupling between Cu1 and Cu2 is weak. Recall that excess O introduces holes, and thus the Fermi level of $Ba_2CuO_{3.3}$ is greatly shifted downward in comparison with $Ba_2CuO_3$, approaching the O-dominated valence bands. If we focus on the energy range around the flat band, the DFT bands can be fitted by a tight-binding model with one single orbital on each Cu [Fig. 4(b)]:

$$H_{hop} = \Delta \sum_{i \in \{Cu2\}} c_i^\dagger c_i + t_\parallel \sum_{<i,j> \in \{Cu1\}} (c_i^\dagger c_j + c_j^\dagger c_i) + t_\perp \sum_{\substack{<i,j>, \\ i \in \{Cu1\}, \\ j \in \{Cu2\}}} (c_i^\dagger c_j + c_j^\dagger c_i) \qquad (2)$$

The first term reflects a difference of the onsite energy between Cu1 and Cu2 due to the different crystal field. The second term is hopping along the chain leg, and the last term is hopping between Cu1 and Cu2. The fitting parameters are given under Fig. 4. We note that $t_\perp$ is indeed one order of magnitude smaller than $t_\parallel$.

What is the nature of the orbitals associated with the hopping model? Figure 4(c) shows the norm of wavefunction at three typical k-points and bands. For the flat band arising from Cu2, the wavefunction can be regarded as a dominating Cu $d_{y2-z2}$ orbital component hybridized with the surrounding O p-orbitals forming σ bonds. This result can be naturally explained by the local square $CuO_4$ geometry in the *yz* plane according to the crystal field theory. For the dispersive bands from Cu1, the orbital component varies from the band top to the band bottom. At high energy, the two dispersive bands are nearly degenerate, and the wavefunction takes the form of Cu $d_{x2-z2}$ oribtal, which predominantly bonds with O along the Cu-O chain. When approaching the O valence bands, the wavefunction rotates to the *yz* plane, and significant bonding with the O atoms shared by Cu2 can be observed, leading to energy splitting of the two dispersive bands. One important implication is that Cu1 and Cu2 can still have strong superexchange coupling。

In summary, we predict a unique ladder structure in the new high-$T_c$ cuprate superconductor Ba$_2$CuO$_{4-\delta}$, by searching over 1800 vacancy structures using the cluster expansion method. In contrast to CuO chains, this ladder structure is not known to exist in other quasi-1D cuprates, which may closely relate to the observed 73K superconductivity. Analysis within the DFT framework indicates that carriers will mainly hop along the two legs of the ladder, while the middle Cu can still couple to two neighboring Cu chains through the superexchange coupling via the shared O atoms.

It is interesting to note that a much enhanced string-like pairing between the doped holes can be explicitly shown [19] (cf. Appendix B3 in Ref. [19] ) in a t-J chain that is coupled to a neighboring t-J chain by superexchange J only without hopping (i.e., $t_\perp = 0$), which thus implies a strong pairing in the present ladder structure.[20] A detailed analytic and density-matrix-renormalization-group (DMRG) study [21], based on the ladder structure shown in Fig. 3(b), will be presented elsewhere.

**Method**

All the first-principle density functional theory calculations are carried out by the projector-augment wave method as implemented in the Vienna *Ab initio* Simulation Package (VASP) using Perfew-Burke-Ernzerhof (PBE) exchange-correlation functional [22]. The energy cutoff for plane-wave basis is set to be 520eV. The electronic convergence criterion is $10^{-6}$eV. A Γ–centered k-point mesh is generated automatically to achieve a sampling density of around 2000 k-points per reciprocal atom for all the structures. All the structures are fully relaxed until the force on each atom is less than 0.01 eV/ Å. Both the CE and Monte Carlo simulation are performed by using the Alloy Theoretic Automated Toolkit [23]. There are two $CuO_2$ layers in a $Ba_2CuO_4$ conventional cell. For the purpose of CE, we consider a quasi-2D $Ba_2CuO_4$ structure containing only one $CuO_2$ layer.

We note that the nominal valence of Cu deviates from +2 in $Ba_2CuO_{4-\delta}$, unless $\delta = 1$. For a doped Mott insulator, the +U correction is not justified. All the results shown in the present work are obtained by PBE without +U correction. With respect to the total energy, PBE is still a good approximation as benchmarked in Tab. I with the experimental structural parameters at both $\delta = 0$ and $\delta = 1$.

**Acknowledgments:**

Z.W. acknowledges the helpful discussions with X.-H. Chen and H.-C. Jiang. This work is supported by NSFC under Grant Nos. 11774196 and 11534007, MOST of China under Grant Nos. 2015CB921000 and 2017YFA0302902, and Tsinghua University Initiative Scientific Research Program.

**Author contributions:**

Z. W. conceived the project. Z. L. designed the calculation. Y. L. and S. D. performed the calculations together. All the authors discussed the results and made contribution to the manuscript.

**Competing interests:**

The authors declare that they have no competing interests.

**Tables and Figures**

Table 1. A comparison of the calculated and experimental $Ba_2CuO_3$ and $Ba_2CuO_4$ structural parameters.

| $Ba_2CuO_3$ | a/Å | b/Å | c/Å | $d_{Cu-O}$(in-pane)/Å | $d_{Cu-O}$(apical)/Å |
|---|---|---|---|---|---|
| Cal. | 3.86 | 4.18 | 13.2 | 1.95 | 2.04 |
| Exp.[8] | 3.9 | 4.1 | 12.9 | 1.99563 | 2.05000 |
| $Ba_2CuO_4$ | a/Å | b/Å | c/Å | $d_{Cu-O}$(in-pane)/Å | $d_{Cu-O}$(apical)/Å |
| Cal. | 4.08 | 4.08 | 13.10 | 2.00 | 1.94 |
| Exp.[7] | 4.00 | 4.00 | 12.94 | 2.0015(2) | 1.861(8) |

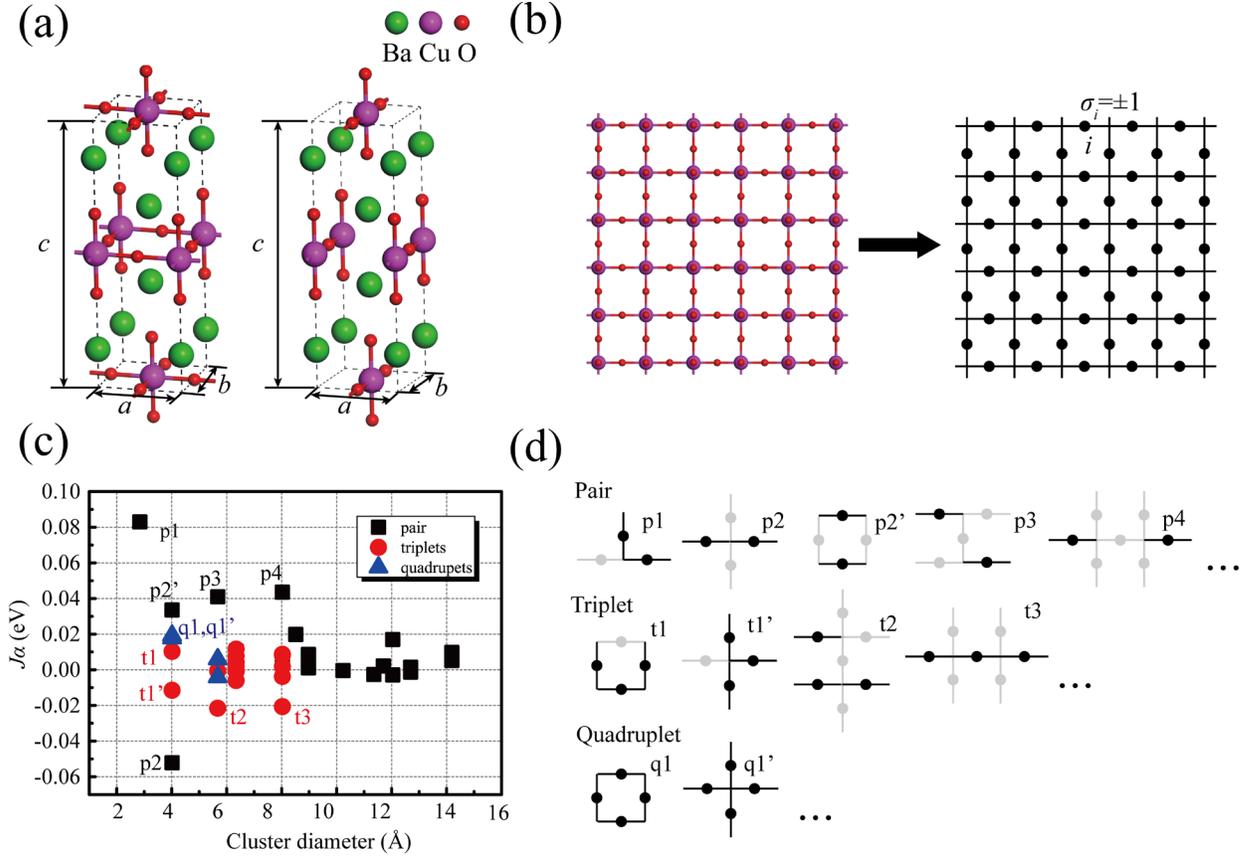

**Fig. 1**(a) Atomic structures of $Ba_2CuO_4$ and $Ba_2CuO_3$ compounds, in which the unit cells are enclosed by the dash lines. The green, purple and red colors represent Ba, Cu, O atoms, respectively. (b) The presence of in-plane O vacancies in the $CuO_2$ plane can be mapped to an Ising-like lattice. Each solid circle represents an oxygen site. (c) Effective cluster interactions ($J_\alpha$) of pair, triplet, quadruplet interactions as a function of cluster diameter. (d) Representative clusters self-learned by the automated CE algorithm, with their strength marked in (c). The solid circles are the O sites involved in the corresponding $J_\alpha$. The shaded circles and lines serve as a guide for eyes of the lattice.

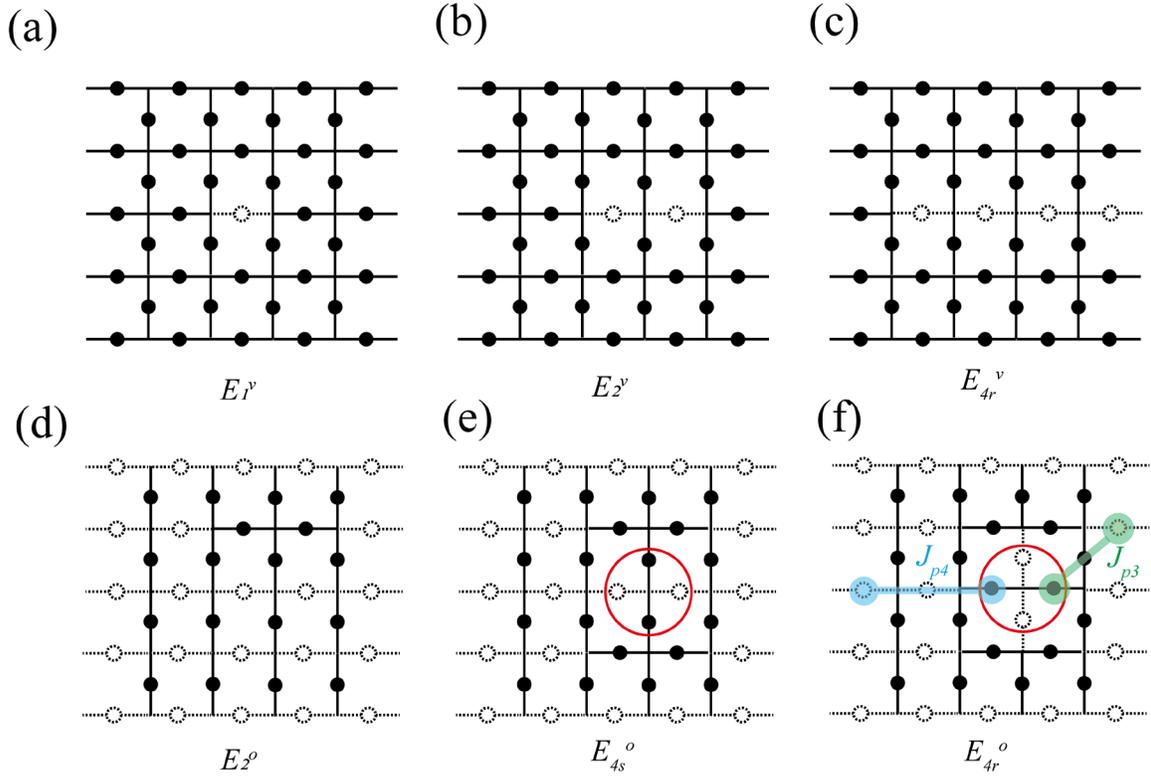

**Figs. 2**(a-c) O vacancy structures used for an analytical discussion around $\delta = 0$. The (d-f) excess O structures used for an analytical discussion around $\delta = 1$. The solid circles are sites that are occupied by O atoms and the dash circles are the empty sites.

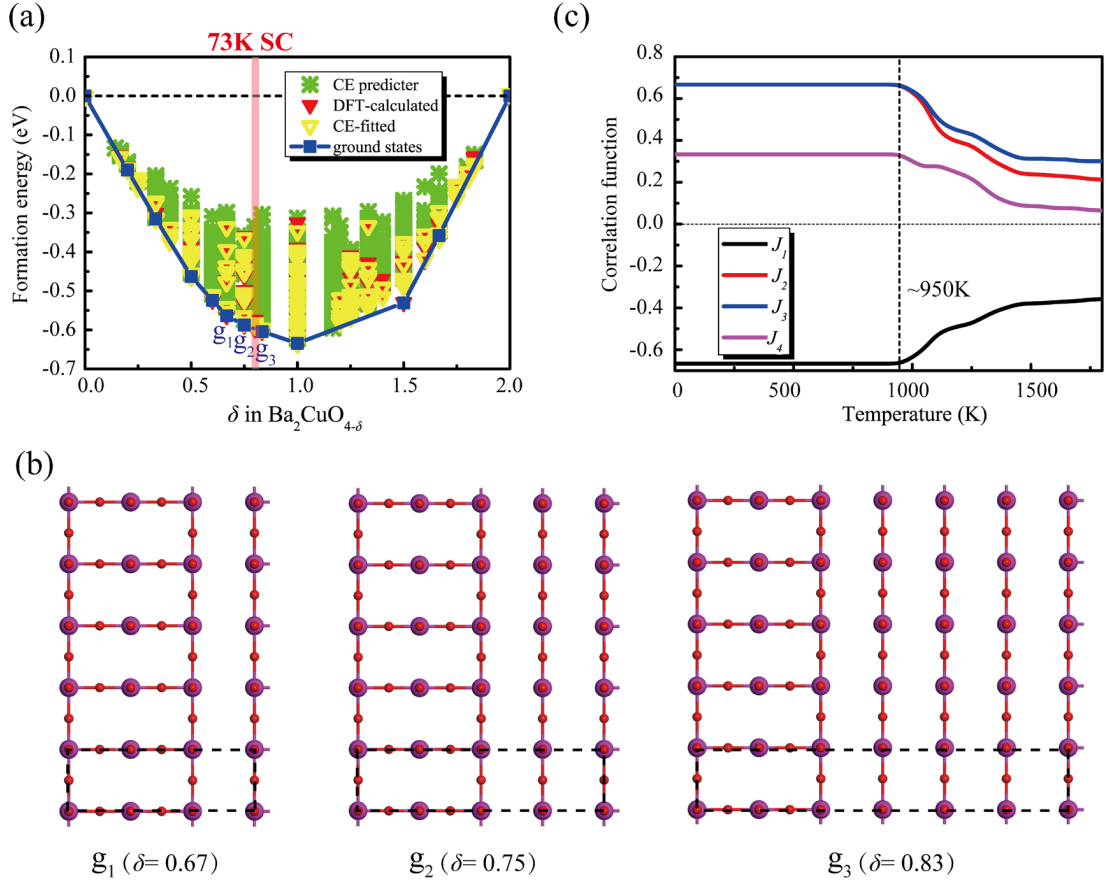

**Fig. 3** (a) Energy of 150 O-vacancy structures calculated by DFT (red triangle) and fitted by CE (yellow triangle) and CE predicted energy of 1813 O-vacancy structures (green star). The lowest energy of different vacancy concentrations predicted by CE and confirmed by DFT are marked by blue rectangles. (b) Lowest-energy structures at $\delta$=0.67, 0.75, 0.83 in $Ba_2CuO_{4-\delta}$. The unit cells are enclosed by the dash-lines. (c) Pair correlation function $C_\alpha = \langle \prod_{i \in \alpha} \sigma_i \rangle$ at $\delta$=0.83 as a function of temperature calculated from MC, which reflects the thermodynamic stability of the ordered structure as shown in (b).

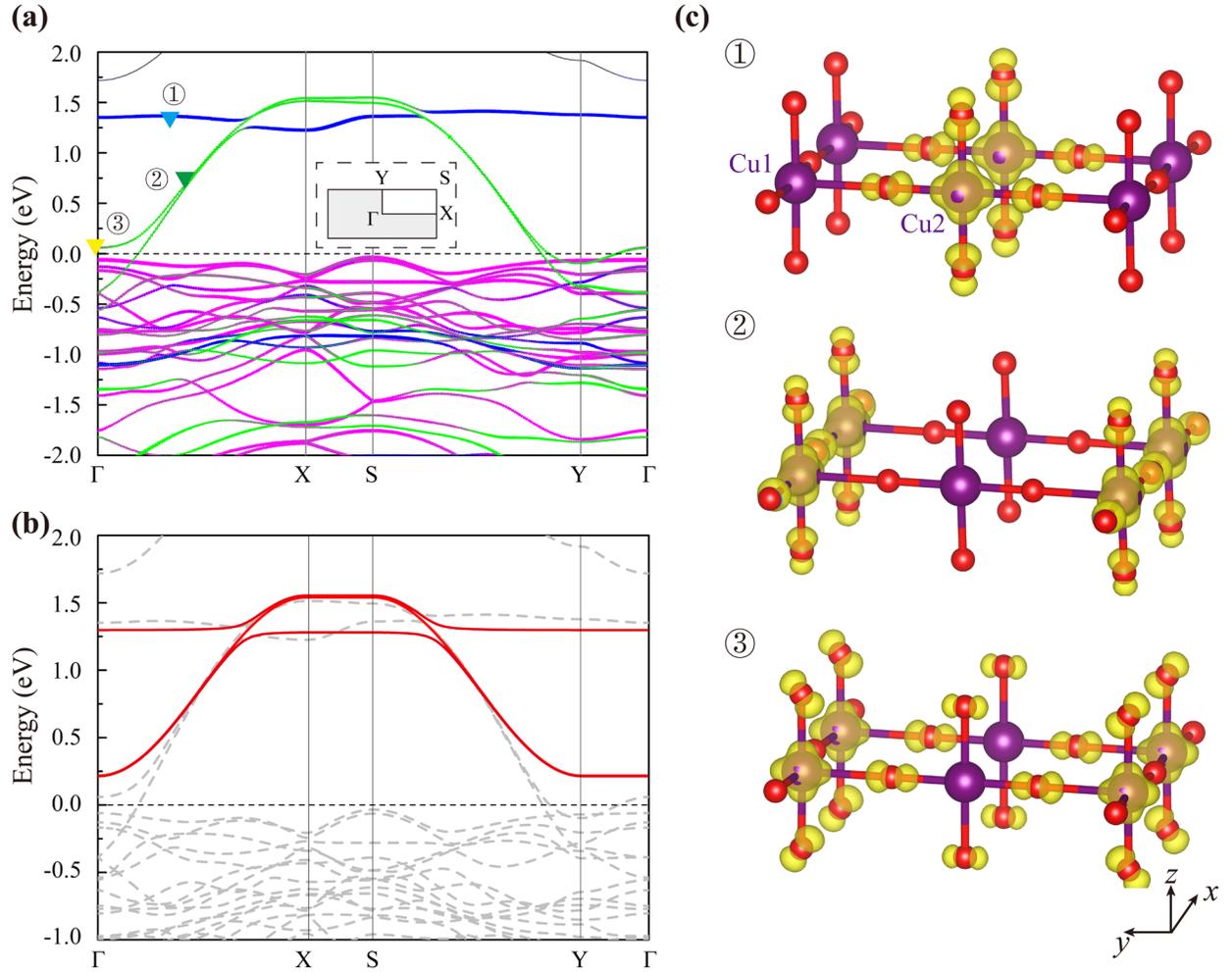

**Fig. 4** (a) DFT band structure of the $\delta$=0.67 lowest-energy structure as shown in Fig. 3(b). The green, blue and peak colors represent $d$ orbitals of Cu1(the Cu atom that is on the ladder legs) and Cu2 (the Cu atom that is in the middle) and $p$ orbitals of O atoms, respectively. The first Brillouin Zone (BZ) is shown in the inset. (b) Fitted band structure (red solid) from tight-binding (TB) model with one single orbital on each Cu. The grey dashed curves are the DFT bands as shown in (a). The fitting parameters are $\Delta = 0.42$ eV, $t_\parallel=-0.33$ eV and $t_\perp = 0.04$ eV. (c) Charge densities of different Bloch states as marked in (a).